\documentclass[aps,preprint,showkeys]{revtex4-2}
\usepackage{graphicx}
\usepackage{amsmath}
\usepackage{makeidx}
\usepackage{amsfonts}
\usepackage{amssymb}
\usepackage{mathtools}
\usepackage{xcolor}
\usepackage{tabularx}

\begin{document}

\title{Messenger size optimality in cellular communications}
\date{\today}

\author{ Arash Tirandaz $^{\rm a,b,\dagger}$, Abolfazl Ramezanpour$^{\rm a,c,\dagger}$, Vivi Rottschäfer$^{\rm d,e}$, Mehrad Babaei$^{\rm a}$, Andrei Zinovyev $^{\rm f,g}$, Alireza Mashaghi$^{\rm a}$$^{\ast}$, \\\vspace{6pt} $^{a}$
{\em{Medical Systems Biophysics and Bioengineering, Leiden Academic Center for Drug Research, Faculty of Science, Leiden University, Leiden, The Netherlands}}; \\$^{b}${\em{Department of Physical Chemistry, Faculty of Chemistry and Petroleum Sciences, Bu-Ali Sina University, 
Hamedan, Iran}};\\ $^{c}$
{\em{Department of Physics, College of Science, Shiraz University, Shiraz, Iran}};\\ $^{d}$
{\em{Mathematical Institute, Faculty of Science, Leiden University, Leiden, The Netherlands}};\\$^{e}${\em{Korteweg-de Vries Institute for Mathematics, University of Amsterdam, Amsterdam, The Netherlands}};\\$^{f}${\em{Institut Curie, INSEM U900, PSL Research University, Paris, France}};\\$^{g}${\em{EvoTec, In Silico Research and Development, Toulouse, France}}\\$^{\dagger}${\em{Equal contributions}}
\\\thanks{$^\ast$Corresponding author. Email: a.mashaghi.tabari@lacdr.leidenuniv.nl
\vspace{6pt}}
\received{6jul 2017} }

\begin{abstract}
Living cells presumably employ optimized information transfer methods, enabling efficient communication even in noisy environments. As expected, the efficiency of chemical communications between cells depends on the properties of the molecular messenger. Evidence suggests that proteins from narrow ranges of molecular masses have been naturally selected to mediate cellular communications, yet the underlying communication design principles are not understood. Using a simple physical model that considers the cost of chemical synthesis, diffusion, molecular binding, and degradation, we show that optimal mass values exist that ensure efficient communication of various types of signals. Our findings provide insights into the design principles of biological communications and can be used to engineer chemically communicating biomimetic systems.

\end{abstract}


\maketitle
\section{Introduction}\label{S0}
Cell-to-cell communications underlie the biology of multicellular organisms and are critical for their function and homeostasis. Organs and tissues are typically made of highly interacting cells that send and receive signals via direct contacts, paracrine or endocrine routes. Communications are not only important for resident cells but also mediates the trafficking of cells into and out of tissues. The latter is typically mediated by secretory factors, known as chemokines, whose gradient is sensed by target migratory cells. In higher vertebrates like humans, sensing chemokines is key to many vital processes such as the recruitment of immune cells to the site of injury \cite{1,2,3}. Despite significant progress in understanding cellular communication processes, their physical design principles remain unresolved.\\

\begin{figure}
\includegraphics[width=12cm]{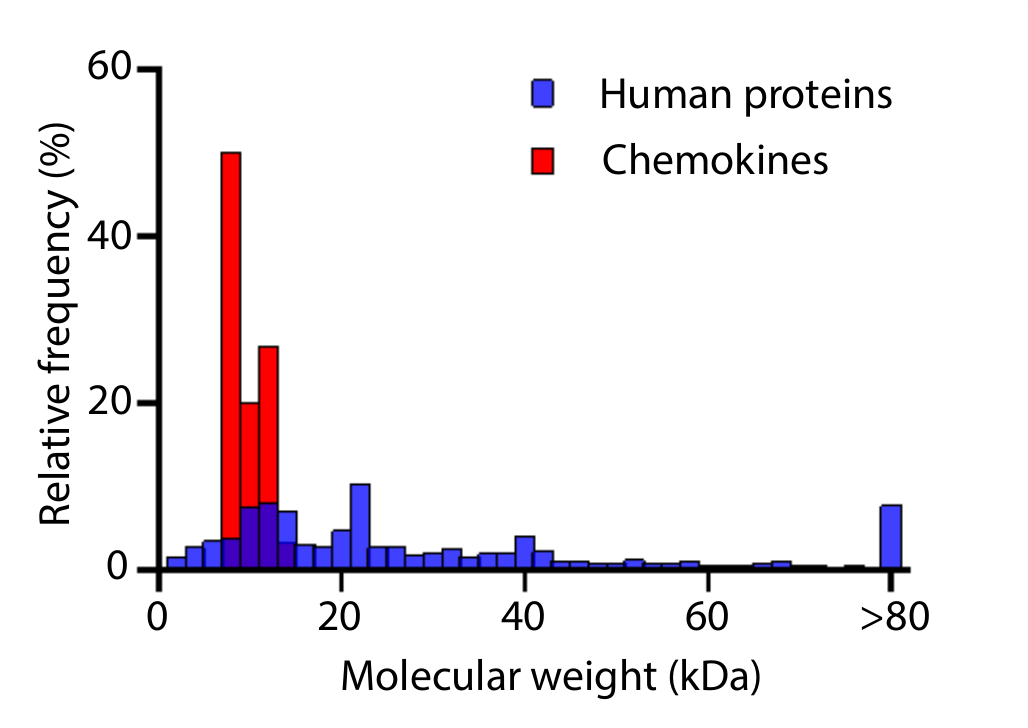}
\caption{The histogram shows that although proteins exist in a wide range of molecular weights from around 0.26 to 4000 kDa, the molecular weights of human chemokines are confined to a narrow range of approximately 8 to 14 kDa. This data, derived from the UniProt database, includes the molecular weights of 204,088 human proteins \cite{uniprot}.}\label{f0}
\end{figure}

A fundamental question in designing a communication strategy is the choice of an optimal carrier (or messenger), which is typically a secreted molecule. An efficient signaling strategy would use minimum energy to send a desired signal within a desired time frame. A carrier molecule may be chosen to have a wide range of physicochemical properties, such as size or molecular mass. The size of a carrier determines the production rate and diffusion coefficient, thus could constrain the information carrying capacity \cite{4}. Remarkably, specialized signal carriers such as a large variety of chemokines have molecular masses that lie within a very limited range of magnitudes, i.e., between 8-10 kDa \cite{5, 6}. It seems that signaling machinery used in cell-to-cell communications are optimized for a specific interval of mass and size of the signal carriers. Figure \ref{f0} displays the data derived from the UniProt database, which includes the molecular weights of 204,088 human proteins \cite{uniprot}.\\

In this study, we investigate how messenger size optimization arises in cellular communications. We take a simplified approach to identify the generic physical mechanisms that lead to possible optimality. Several key steps in cellular communications are affected by carrier size. The energy expenditure for synthesizing proteins correlates with size, and larger proteins diffuse slower in the medium \cite{101,102}. Extracellular protein degradation is a complex process influenced by numerous factors, including protein size, which in turn affects protein lifetime and binding affinity in intricate ways. Despite this complexity and limited experimental data, simplified trends can be observed. Although there is no strict rule, smaller proteins are often more susceptible to degradation due to their simpler structures, fewer stabilizing interactions, and higher rates of renal filtration. It has been observed that renal elimination decreases with increasing molecular size \cite{117}. Meanwhile, the extracellular proteasome, the primary catalyst for enzymatic degradation of extracellular proteins, exhibits lower degradation rates for larger proteins \cite{130, 21}. For instance, smaller extracellular proteins, such as cytokines and growth factors, generally have shorter lifetimes than larger extracellular proteins, such as matrix proteins and immunoglobulins \cite{115}. Larger proteins tend to have more disulfide bonds, which further stabilizes them against degradation; this trend generally applies to proteins or polypeptides longer than 50 residues \cite{116}. Glycosylation also plays a role in protein stability, with larger proteins exhibiting a greater degree of glycosylation due to their larger surface areas in folded structures. Protein glycosylation is well-established to impact protein stability \cite{118}. Importantly, protein size influences ubiquitination processes related to proteasomal degradation. Larger proteins require extended ubiquitination chains, while smaller proteins (typically those with fewer than 150 residues) are often degraded following monoubiquitination \cite{119,120}. Overall, as argued above, larger proteins tend to be more stable, while smaller proteins exhibit less stability and greater flexibility. This increased flexibility is associated with higher conformational entropy, which can influence binding affinity to receptors \cite{121}. Finally, the shape (concentration time dependence) of the initial signal is possibly another critical factor in designing a communication strategy. Various signal shapes are seen in biology, including step-like profiles \cite{7}, pulse-like profiles \cite{8}, and  oscillatory signals \cite{9,10,11}. Various time-dependent and independent communication strategies can be seen in cell biology \cite{12,FFL-sr-2018,SLL-jms-2019,EBV-prl-2020,BM-prl-2023,PJ-ieee-2016,SAC-prr-2023,B-nri-2023}. Here, we focus on proteins and polypeptides as signal carriers or messengers, and examine the occupancy of receptors on the receiving cell and its dependency on the messenger's molecular size (mass) and spatio-temporal characteristics of the input signal.\\

\section{Basics of chemokine-mediated cellular communications}\label{S1}
In this section, we briefly describe the main elements and characteristics of chemokine-mediated communication processes which we shall use in the next section for a numerical simulation of the cellular communication. In particular, we discuss the dependency of each step in a communication process on the size/mass of the messenger which carries the signal. 
   
Let us assume that the transmitter cell sends messengers $P$ to the receiver cell located at a distance of $L_d$. In the case of chemokine messengers, $P$ is a protein produced inside a transmitter, and it diffuses through the environment and binds to a receptor on the surface of the receiver. $P$ can be degraded both inside and outside the transmitter; however, extracellular degradation is the dominant process, as $P$ typically spends most of its lifetime in the extracellular space. The probability of the receptor being occupied represents the cells' ability to respond effectively to the concentration of ligand molecules $P$. More precisely, we use $P_t(\mathbf{r})$ for the number density of proteins at position $\mathbf{r}$ and time $t$. 

The protein is produced inside the transmitter cell with the following synthesis rate (see Supplementary Information)\cite{Gorban} 
\begin{align}\label{fs}
\frac{dP}{dt}|_{syn}=K(\alpha_0+\alpha_1e^{-k_d t}-\alpha_2e^{-(k_3+k_{Rd}+k_P) t}+\alpha_3e^{-(k_1+k_d+k_2) t})=f_s(t).
\end{align}
The parameters $K$ and $\alpha_0,\alpha_1,\alpha_2,\alpha_3$ depend on the reaction rates $k_1,k_2,k_3,k_d,k_P,k_{Rd}$, and another reaction rate $k_T$, which are defined more precisely in the Supplementary Information. Here, $k_{1}$ is the rate constant of translation initiation, $k_{3}$ is the rate of
nullifying active ribosomes, and $k_{P}$ represents the rate constant of protein synthesis \cite{Bostrom, Siwiak, Sharma}. These three rate constants $k_{1}$, $k_{3}$, and $k_{P}$ directly link the size of the protein to the kinetics of its synthesis. We consider $k_1,k_3,k_P \propto L_p^{-\lambda}$, which represents the proportionality of rate constants with the length of the protein $L_p$ (e.g., the number of amino acids) and $\lambda$ is a tunable positive real number. More precisely, we consider the following reaction rates
\begin{align}
k_1&=k_3=k_P=\frac{1}{\tau_k}\frac{1}{L_p^{\lambda}},\\
k_2&=k_{Rd}=k_T=\frac{1}{\tau_k},\\
k_d&=\frac{4}{\tau_k},
\end{align}
characterized by the time scale $\tau_k$. 

Assuming an Arrhenius-type rate constant, one can show that at steady-state, the overall activation energy of synthesis is an inverse function of the length of the protein. The cost of synthesis is reflected in the activation energy required for protein synthesis, which directly depends on the logarithm of the synthesis rate \cite{18},
\begin{align}\label{Ks}
K_s=\frac{k_{T}k_{2}L_p^{-2\lambda}(L_p^{-\lambda}+k_{2})}{k_{d}(L_p^{-\lambda}+k_{2})(L_p^{-\lambda}+k_{2}+k_{d})(2L_p^{-\lambda}+k_{Rd})}. 
\end{align}
The energy cost then is given by
\begin{align}\label{DE}
\Delta E_s=-\ln K_s.
\end{align}
In the following, we measure the energies in units of $RT$, where $R$ is the gas constant and $T$ is absolute temperature.
From Eq. (\ref{Ks}) we see that the energy cost of synthesis increases with $L_p$ and $\lambda$. In addition, the diffusion coefficient and the binding probability are decreasing functions of $L_p$. Therefore, to enhance the chance of observing large proteins in the receptors, we should work with small values of $\lambda \simeq 1/4$. The value of $\tau_k$ is then chosen such that for $L_p=1$ we obtain $f_s(0)=1$ and $\Delta E_s\simeq 0$. This fixes the scale of the synthesis rate and energy by demanding to have reasonable values for small proteins.

After production inside the transmitter the protein diffuses and is eventually degraded outside (or inside) the transmitter. Assuming an approximately spherical shape for the messenger protein, the diffusion constant $D$ is given by $ 8.34\times10^{-8} \frac{T}{\eta} M^{-0.33}$ $cm^{2}s^{-1}$, where $M, T, \eta $ are the molecular mass of the protein, absolute temperature, and viscosity of the environment. Since $M\propto L_p$, this means that $D\propto L_p^{-0.33}$.

Next, we discuss the dependence of the degradation reaction on molecular mass. A common mechanism for the degradation of extracellular proteins involves enzyme-substrate reactions, which typically follow Michaelis-Menten kinetics \cite{19, 20, 21, 22, 23, 24, 25}. To model the size-dependency of the degradation parameter, we follow the approach taken in Ref. \cite{26}. The rate of degradation reads as follows
\begin{align}\label{PDeg}
\frac{dP}{dt}|_{deg}=-\frac{V_{max}}{K_M+P_t(\mathbf{r})}P_t(\mathbf{r}).
\end{align} 
Both $K_{M}$ and $V_{max}$, and thus the degradation rate are functions of protein length $L_p$. This aligns with previously reported size-based degradation analyses conducted in various contexts, incorporating both intrinsic and extrinsic factors \cite{26,27,29,30,31,32}.

Finally, we consider the reception of the transmitted signal by the receiver cell, which is mediated by the binding of the messenger protein to the receptors at the receiver cell. We assume that the binding sites are independent of each other, and each binding site can be occupied by only one protein. The association and dissociation constants, $k_{on}$ and $k_{off}$, respectively, of the binding of a protein to the receptor can be related to each other by the equilibrium constant $K_{eq}=k_{on}/k_{off}=e^{-\Delta E_b}$. Here, the binding energy is given by \cite{44}:
\begin{equation}\label{Eb}
\Delta E_b=\Delta E_{b}^{0}+1.5 \ln \frac{\Lambda}{\Lambda_{0}},
\end{equation}
where $\Lambda$ is the effective force constant which directly depends on the mass of the protein $M\propto L_p$. Thus from the above equation we have $k_{on}/k_{off} \propto L_p^{-3/2}$. Here $\Delta E_{b}^{0} $ is the activation energy irrespective of weak forces between protein and receptor. $\Lambda_{0}$ is employed to render the logarithm dimensionless and energies are scaled with $RT$. The flexibility of proteins can be related to force constants. Smaller molecules with fewer values of force constant are usually more flexible and bind to the receptor more tightly \cite{121, 45}.

\section{Numerical Simulation of the Model}\label{S2}
In this section, we use the information provided to simulate a model of transmission of various signals between cells. The three stages of synthesis, diffusion/degradation, and binding/unbinding in the signaling process are simulated as a discrete-time stochastic process of a system of diffusing and non-interacting proteins. The neighboring cells are represented by the surface of an enclosing sphere, which defines the boundary of our system, with the transmitter source located at the center of this sphere. For simplicity, it is also assumed that the probability of finding a receptor is uniformly distributed on the surface of the sphere. This means that there are nonzero probabilities for collision, binding, and unbinding of the proteins with the receptors at any point of the boundary of the system.  The main results which are presented here are obtained by numerical simulation of a dynamical model which is described in more detail below.

We consider a three dimensional sphere of diameter $2L_d$ with volume $V_d$ and surface of area $A_d$. We denote the 
radius of this sphere by $r=L_d$ and its center by $\mathbf{r}=0$. A sphere with diameter $2L_c$ and volume $V_c$ is placed at the center of the sphere to represent the transmitter cell. We take $L_d$ as the typical distance between the cells which is about $20L_c$. We consider a discrete time process of signaling by a simple diffusion of proteins of length $L_p$ (in terms of an appropriate length scale of the protein) which can occupy any position $(x,y,z)$ in the sphere of volume $V_d$. There is no interaction between the proteins. The number of proteins inside a unit of volume at point $\mathbf{r}$ and time step $t$ is denoted by $P_t(\mathbf{r})$. The signaling process starts at time step $t=0$ with no proteins in the system. At each time step $\rho_g(t)V_c$ proteins are generated uniformly inside the transmitter cell. Here $\rho_g(t)=f_s(t)\phi(t)$ where $f_s(t)$ is the rate of synthesis given in Eq. (\ref{fs}) and $\phi(t)$ with a characteristic time $\tau_s \propto L_d^2$ determines the shape of the output signal of the transmitter. All times in the following are taken proportional to $L_d^2$ which is expected from a standard Brownian motion. The energy cost of generating a single protein is given by $\Delta E_s$ defined in Eq. \ref{DE}. All quantities here are dimensionless depending on the appropriate length/time/energy scales of the system. The time interval between two successive time steps is taken $\Delta t=1$

\begin{figure}
\includegraphics[width=12cm]{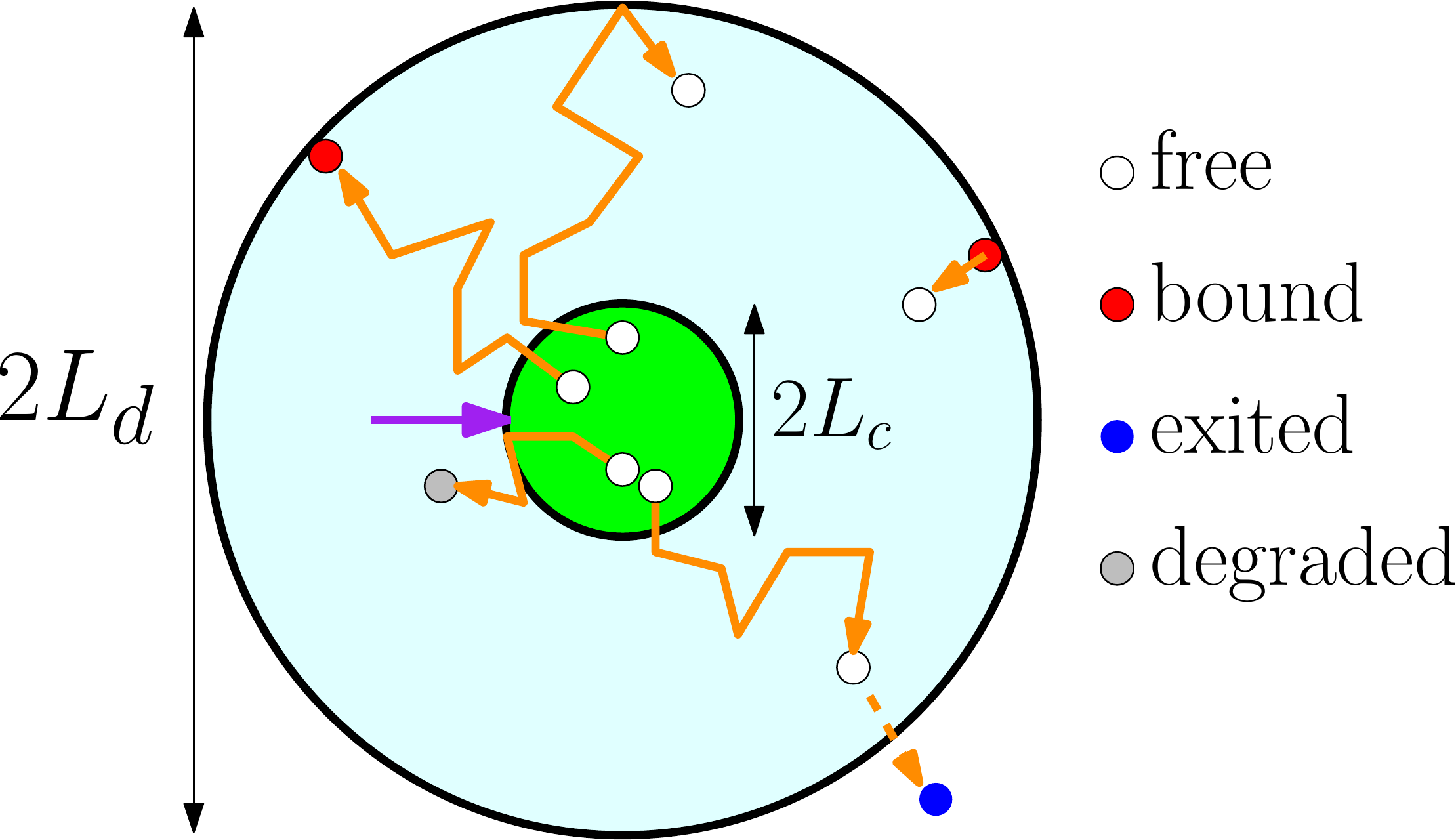} 
\caption{A schematic of cell communication by diffusion within a three dimensional sphere. The large filled circle shows a two-dimensional cross section of the volume enclosed by the sphere. Proteins of length $L_p$ are generated in the central cell (in green) of volume $V_c=(4/3)\pi(L_c)^3$. A free protein diffuses and is degraded during the motion. It can reflect from or bind to the boundary surface in case of collision with the neighboring cells. The collision with the neighboring cells is effectively modeled by a probability of collision $p_{coll}$. A free protein diffuses and is degraded during the motion. It can reflect from or bind to the boundary surface in case of collision with the neighboring cells. A free protein can exit the system forever if no collision with the other cells occurs, while a bound protein can be released as a free protein within the enclosing sphere.}\label{f1}
\end{figure}

Figure \ref{f1} displays an illustration of the process. A protein can move freely inside the sphere (free), bind to a neighboring cell (bound) on the boundary, reflect from the surface as a free protein inside the sphere, or exit from the system (exited). We assume that an exited particle never returns to the system inside the sphere of volume $V_d$. A bound protein has a nonzero chance to become a free protein with probability $p_{off}\propto k_{off}\Delta t$. The total number of generated proteins and the total energy cost up to time step $t$ are denoted by $N(t)$ and $E(t)$, respectively. The number of free and bound proteins at time step $t$ are shown by $F(t)$ and $B(t)$, respectively.  

More precisely, at any time step $t<T$ we do the following to update the state of the system in $T\propto L_d^2$ time steps:
 
\begin{itemize}

\item A free protein changes its position to $(x+\Delta x,y+\Delta y,z+\Delta z)$ where $\Delta x, \Delta y, \Delta z$ are Gaussian random variables of mean zero and variance $D$, where the diffusion coefficient $D$ is given in the previous section.

\begin{enumerate}

\item If the protein is still inside the sphere, i.e., its distance $r=\sqrt{x^2+y^2+z^2}$ to the center of the sphere is less than $L_d-\Delta L$, it can be degraded with probability $p_{deg}$. Here $\Delta L \simeq L_c$ is to consider the effects of the volume occupied by the neighboring cells at the surface of the sphere.

\item If the protein is close to the boundary of the sphere, i.e., $r$ is larger than $L_d-\Delta L$ and less than $L_d$, the protein collides with a neighboring cell with probability $p_{coll}\propto A_c/A_d$. Otherwise, it exits from the system and never returns to it again. Here $A_c=4\pi L_c^2$ is the effective cross section of a neighboring cell.    
 
\item In case of collision, the protein binds to the neighboring cell with probability $p_{B}=p_{on}\max\{1-B(t)\frac{A_p}{A_c},0\}$, where $p_{on} \propto k_{on}\Delta t$. Otherwise it is reflected and returns to the sphere as a free protein within distance $\Delta L$ from the boundary surface. Here $A_p$ is the effective surface occupied by a bound protein of volume $V_p$. By choosing $\max\{1-B(t)\frac{A_p}{A_c},0\}$ we assume that the number of binding sites on the surface of the sphere is limited. Thus, $p_{on}$ is the binding probability conditioned on the availability of a binding site. In the following, we set $A_p=1$ and $V_p=1$ independent of the protein length; only the ratios $A_p/A_c$ and $V_p/V_d$ appear in the equations, that is $A_c$ and $V_d$ are measured in terms of $A_p$ and $V_p$, respectively. Note that in the reflection and the binding processes the position of the protein does not change by the diffusion process. This means that only one of the above processes can occur at each time step to clearly separate the effects of the processes.   
 
\end{enumerate}

\item A bound protein becomes a free one with probability $p_{off}$ and remains inside the sphere within the distance $\Delta L$ from the enclosing surface. This means that in this process the freed protein does not change its position by the diffusion process. The unbinding probability $p_{off}$ is taken proportional but smaller than the collision probability $p_{coll}$.      

\end{itemize}
   
Note that the effect of the neighboring cells are effectively modeled by introducing a collision probability $p_{coll}$ which is proportional to $A_c$, the surface of a cell of radius $L_c$.    
Some of the model parameters are fixed by the typical empirical/theoretical values found in the literature. For other free parameters, we choose reasonable values to fix the scales and have a trade off in the energy cost, diffusion, and binding probability with the rate of degradation vs the protein length. More precisely, we work with the following model functions and parameters.
Three types of signal shapes are studied
\begin{align}
\phi(t)=\begin{cases}
1 & t\le \tau_s\\ 
0 & t>\tau_s
\end{cases},\label{stp} \\ 
\phi(t)=\exp(-t/\tau_s),\label{exp} \\ 
\phi(t)=\frac{1}{(t/\tau_s+1)^{\gamma}}, \label{pow}
\end{align}
with $\tau_s\propto L_d^2$. The step and exponential signals (\ref{stp}) and (\ref{exp}), respectively, represent two extreme limits from a constant to a rapidly decreasing function. The power-law signal in (\ref{pow}), with an exponent $\gamma\simeq 2$, represents a typical behavior in between.

For the diffusion coefficient and degradation probability of a single protein we take
\begin{align}
D &=\frac{4\pi}{L_p^{0.33}},\\
p_{deg} &=\frac{V_{max}}{K_M+P_t(\mathbf{r})},\\
V_{max} &=\frac{1}{L_d},\hskip0.5cm K_M = (\frac{L_p}{2})^{\delta}.
\end{align}
We assume that the probability of degradation is rapidly decreasing with $L_p$ to compensate the other processes which are suppressing the transmission of larger proteins. In the numerical simulations we choose $\delta\simeq 2$. In addition, we consider also dilution effects by assuming that $V_{max}=1/L_d$ such that a single protein of size $L_p=2$ will be degraded with probability $\simeq 1$ in about $L_d$ time steps after its production. For the sake of simplicity, we approximate $P_t(\mathbf{r})$ with the average number of free proteins $F(t)V_p/V_d$ in the volume $V_p$ occupied by a single protein at time step $t$.

Finally, the binding and unbinding coefficients are given by
\begin{align}
p_{on} &= (\frac{L_p}{2})^{-\beta},\\
p_{off} &=\frac{1}{z_{off}}p_{coll},
\end{align}
where we choose $\beta=3/2$ as expected for the ratio $k_{on}/k_{off}$ explained after Eq. (\ref{Eb}). It is also assumed that the unbinding probability is smaller than the collision probability by a factor of $z_{off}=10$. 

The dependence of some parameters on the protein length (e.g., $D\propto L_p^{-0.33}$ and $p_{on}/p_{off}\propto L_p^{-3/2}$) is determined from the arguments of the previous section. For the other parameters, we follow the qualitative behavior described in that section. Note that a nontrivial behavior with $L_p$ is expected only if the lower degradation probability of larger proteins compensates for the other suppressing factors in the signaling process which increase with $L_p$. For instance, the chance of binding a large protein to a receptor at the surface of the sphere decreases monotonically with protein length if $\delta \ll 2$ and $z_{off}\ll 10$, given the above parameters. Finally some parameters like $\tau_k$ are used to set the scale of a quantity like the energy cost.     

\section{Results}\label{S21}
\begin{figure}
\includegraphics[width=16cm]{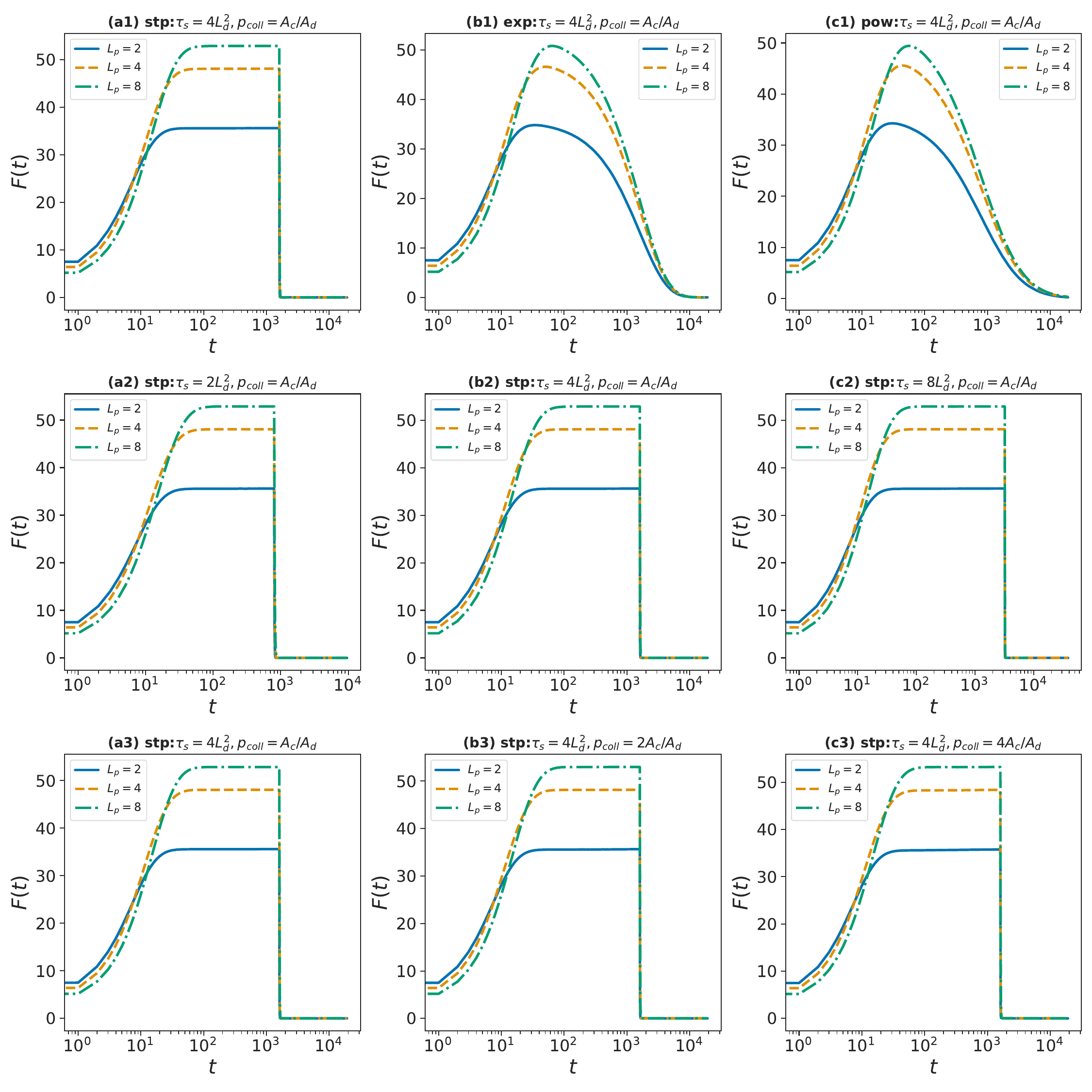} 
\caption{Time dependence of the number of free proteins $F(t)$ on a logscale for $L_p=2$(solid), $L_p=4$(dashed), and $L_p=8$(dash-dot). Here $L_c=1, L_d=20, T=12\tau_s, \tau_k=1/72, \lambda=1/4, \gamma=2, \delta=2, 1/z_{off}=0.1$. Panels ((a1),(b1),(c1)): for the three types of signals $\phi(t)$: step (stp), exponential (exp), and power law (pow) functions with $\tau_s=4L_d^2, p_{coll}=A_c/A_d$. Panels ((a2),(b2),(c2)): for different $\tau_s$ when the signal $\phi(t)$ is a step function with $p_{coll}=A_c/A_d$. The cell and sphere surfaces are $A_{c,d}=4\pi L_{c,d}^2$. Panels ((a3),(b3),(c3)): for different $p_{coll}$ when the signal $\phi(t)$ is a step function with $\tau_s=4L_d^2$. The results are averaged over $10^6$ realizations of the signaling process (synthesis, diffusion and degradation, binding and unbinding).}\label{f2}
\end{figure}

\begin{figure}
\includegraphics[width=16cm]{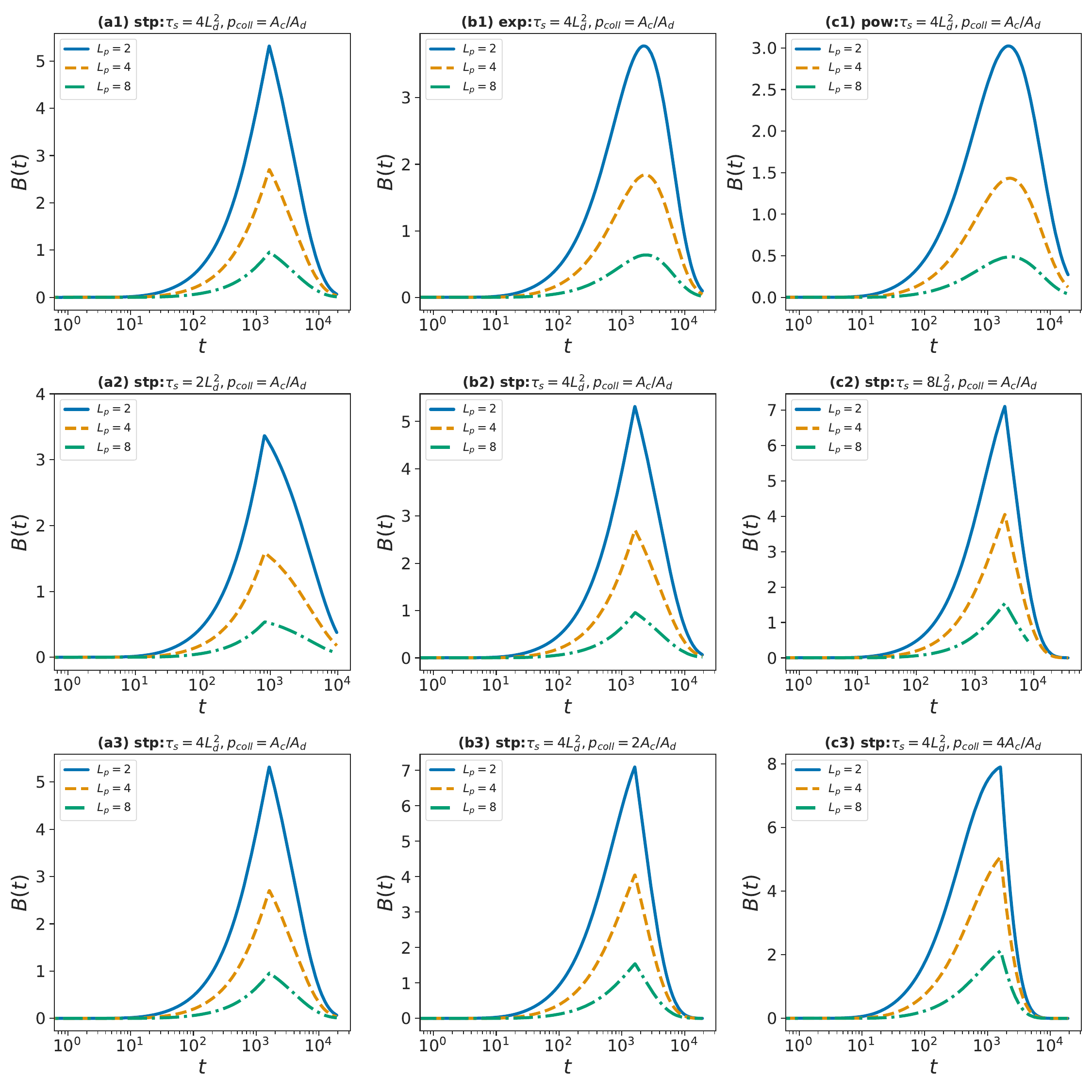} 
\caption{Time dependence of the number of bound proteins $B(t)$. Here $L_c=1, L_d=20, T=12\tau_s, \tau_k=1/72, \lambda=1/4, \gamma=2, \delta=2, 1/z_{off}=0.1$. Panels ((a1),(b1),(c1)): for the three types of signals $\phi(t)$: step (stp), exponential (exp), and power law (pow) functions with $\tau_s=4L_d^2, p_{coll}=A_c/A_d$. Panels ((a2),(b2),(c2)): for different $\tau_s$ when the signal $\phi(t)$ is a step function with $p_{coll}=A_c/A_d$. The cell and sphere surfaces are $A_{c,d}=4\pi L_{c,d}^2$. Panels ((a3),(b3),(c3)): for different $p_{coll}$ when the signal $\phi(t)$ is a step function with $\tau_s=4L_d^2$. The results are averaged over $10^6$ realizations of the signaling process (synthesis, diffusion and degradation, binding and unbinding).}\label{f3}
\end{figure}

Figures \ref{f2} and \ref{f3} ((a1),(b1),(c1)) display the average number of free and bound proteins $F(t)$ and $B(t)$ vs time for the three types of signals and $L_p=2,4,8$. Variations with respect to the signal time $\tau_s$ and collision probability $p_{coll}$ are also reported in the middle and lower rows of the figures for a step signal function. Clearly the exponential and power-law signals display a smoother behavior with time than the step function, as expected. The step signals always display the largest values of free and bound proteins for the same signal time. Moreover, the total number of free proteins shows a plateau before the end of the signal time. The length of this plateau increases with $\tau_s$ but does not depend on the collision probability. As expected, the number of bound proteins grows with both the signal time and collision probability.  

\begin{figure}
\includegraphics[width=16cm]{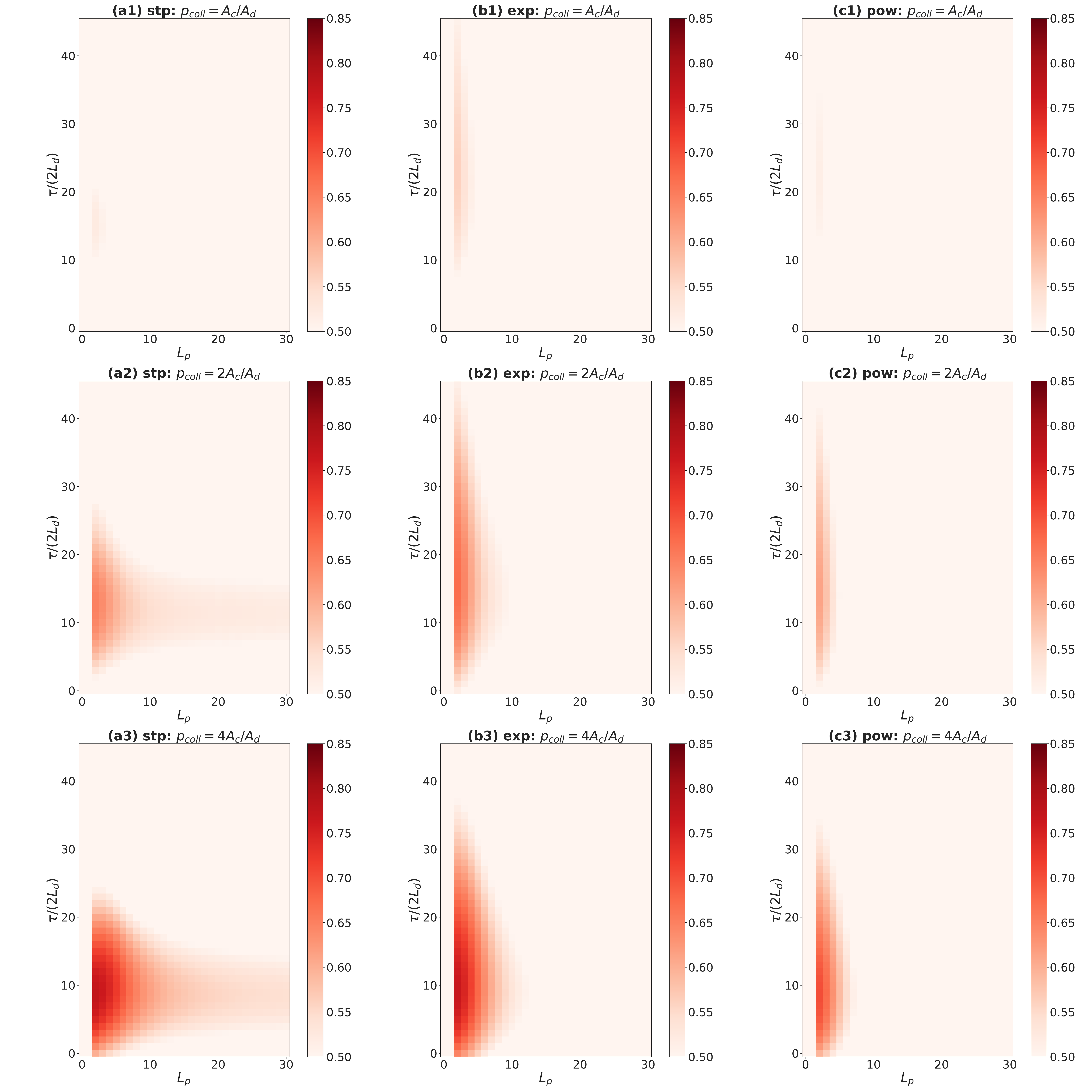} 
\caption{Correlation of the number of the bound and free proteins for time lag $\tau$ and protein length $L_p$. Darker colors correspond to higher correlations. Here $L_c=1, L_d=20, \tau_s=2L_d^2, T=12\tau_s, \tau_k=1/72, \lambda=1/4, \gamma=2, \delta=2, 1/z_{off}=0.1$. The panels show the results for the three types of signals $\phi(t)$: step (stp), exponential (exp), and power law (pow) functions. Panels ((a1),(b1),(c1)): $p_{coll}=A_c/A_d$. Panels ((a2),(b2),(c2)): $p_{coll}=2A_c/A_d$. Panels ((a3),(b3),(c3)): $p_{coll}=4A_c/A_d$. The cell and sphere surfaces are $A_{c,d}=4\pi L_{c,d}^2$. The results are averaged over $10^6$ realizations of the signaling process (synthesis, diffusion and degradation, binding and unbinding).}\label{f4}
\end{figure}

Let us see how much information is provided about the number of bound proteins given the number of free ones for different signal types. To this end, we consider the linear correlation between the two quantities which is statistically less demanding than a quantity like the mutual information. The correlation of the number of bound and free proteins with a time lag $\tau$ is defined as follows, 
\begin{align}
C_{BF}=\frac{\sum_{t=\tau+1}^T\langle B(t)F(t-\tau)\rangle}{\sqrt{\sum_{t=1}^T\langle B(t)^2\rangle}\sqrt{\sum_{t=1}^T\langle F(t)^2\rangle}}.
\end{align}
This measure shows how much the number of bound proteins at time step $t$ is aligned with the number of free proteins at a previous time step $t-\tau$. In Fig. \ref{f4} we report the above quantity vs the protein length $L_p$ and the time lag $\tau$. In general, the correlation is higher for small $L_p$ and shifts to smaller time lags $\tau\propto L_d^2$ (like a Brownian motion) as the collision probability increases.
We also observe that for small protein lengths the correlation $C_{BF}$ is distributed over a larger range of time lags for the exponential and power-law signals than for the step signal. The latter however displays larger correlations for larger $L_p$ compared to the other signals. This means that the exponential and power-law signals are more sensitive to protein lengths regardless of the time lags whereas the step signals display more selectivity in the time lags than the protein lengths.         
Each signal type thus provides a characteristic time-lag and messenger size which result in high correlations between $B(t)$ and $F(t)$.

The main quantities that are studied in the following are:
the first time that a protein binds $T_{min:B}$ and the time at which the number of bound proteins is maximal $T_{max:B}$. 
The number of bound proteins at $T_{min:B}, T_{max:B}$, i.e., $B_{min:B}=B(T_{min}), B_{max:B}=B(T_{max})$ and the total number of proteins $B_{tot}=\sum_{t=1}^T B(t)$ that bind in the time interval $T$. We also study the energy costs at these time points $E_{min:B}=E(T_{min})$, $E_{max:B}=E(T_{max})$, and $E_{tot}=E(T)$ and the number of generated proteins at these times $N_{min:B}=N(T_{min})$, $N_{max:B}=N(T_{max})$, and $N_{tot}=N(T)$. 

Note that the minimal and maximal values are conditioned on the binding of at least one protein. Thus we also define the probability of that event, i.e., $P(B_{tot}>0)$, and report the min/max values conditioned on $B_{tot}>0$. More precisely, it means that a quantity like $B_{min:B}$ is averaged only over realizations with $B_{tot}>0$. The total quantities like $E_{tot}, N_{tot}$ are not restricted by this condition. In this way, the minimum and maximum of the main quantities are obtained from $X_{min/max}=P(B_{tot}>0)X_{min/max:B}$ where $X\in \{T,B,N,E\}$ can be any of the main quantities. Moreover, $B_{tot}(t)$ counts the total number of bound proteins in time interval $T$. A bound protein that remains in the binding site for $n$ time steps is counted $n$ times in this quantity. Therefore, $B_{tot}$ has some information also about the time periods a bound protein remains in the binding site, whereas $B_{min}$, $B_{max}$, and $B(t)$ are giving the number of bound proteins at a given instant of time.

\begin{figure}
\includegraphics[width=16cm]{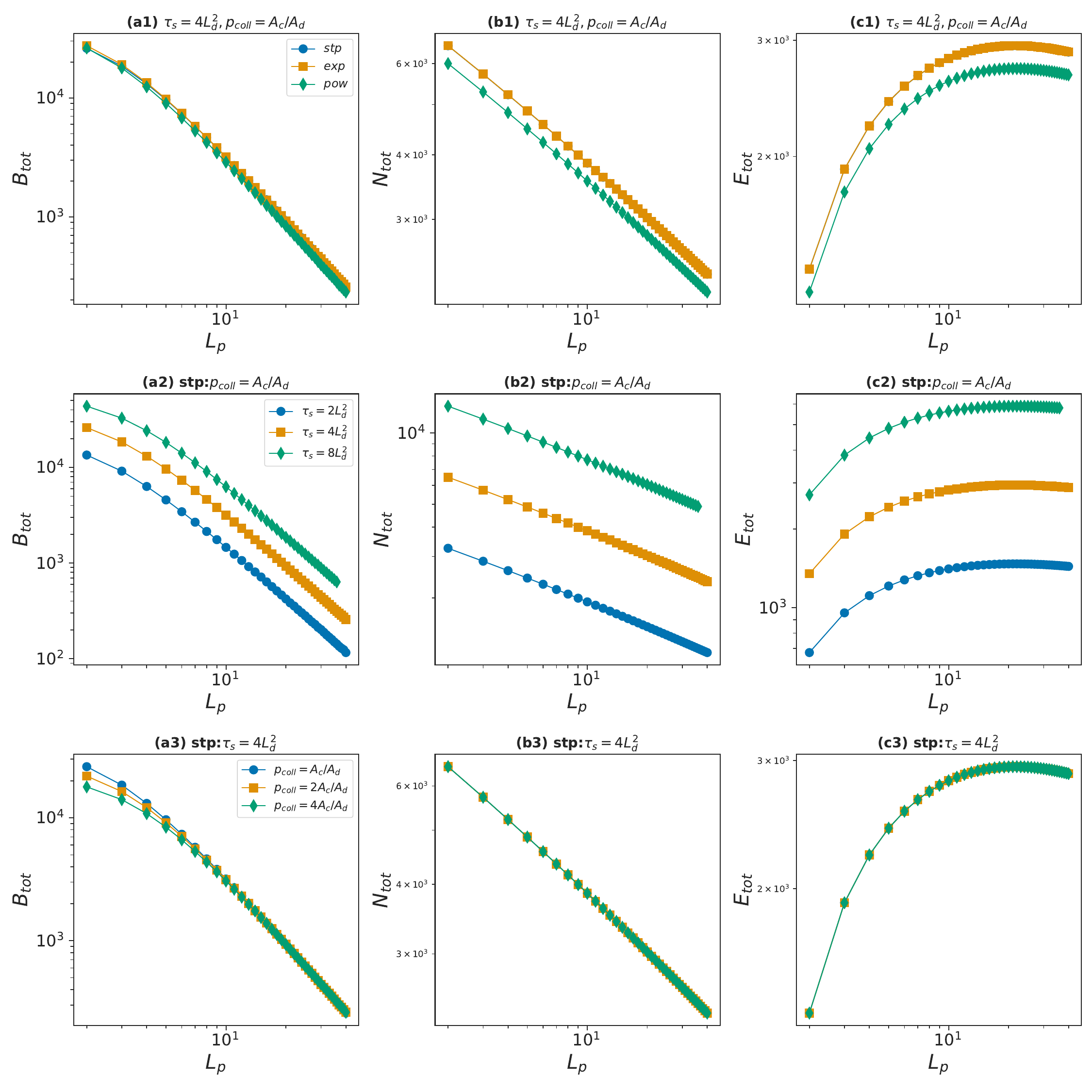} 
\caption{The total of the main quantities $B_{tot},N_{tot},E_{tot}$ vs the protein length $L_p$. Here $L_c=1, L_d=20, T=12\tau_s, \tau_k=1/72, \lambda=1/4, \gamma=2, \delta=2, 1/z_{off}=0.1$. Panels ((a1),(b1),(c1)): for the three types of signals $\phi(t)$: step (stp), exponential (exp), and power law (pow) functions with $\tau_s=4L_d^2, p_{coll}=A_c/A_d$. Panels ((a2),(b2),(c2)): for different $\tau_s$ when the signal $\phi(t)$ is a step function with $p_{coll}=A_c/A_d$. Panels ((a3),(b3),(c3)): for different $p_{coll}$ when the signal $\phi(t)$ is a step function with $\tau_s=4L_d^2$. The cell and sphere surfaces are $A_{c,d}=4\pi L_{c,d}^2$. The legends in each row are the same as the one in the first column. The results are averaged over $10^6$ realizations of the signaling process (synthesis, diffusion and degradation, binding and unbinding).}\label{f5}
\end{figure}

Figure \ref{f5} shows how the total of the main quantities $B_{tot},N_{tot}$ and $E_{tot}$ change with $L_p$ for the three types of signals. We observe that the total numbers of synthesized and bound proteins decrease with the protein length. The energy cost, however, displays a maximum and decays slowly for large protein sizes, if the number of synthesized proteins diminishes more rapidly with $L_p$ than the increase in the energy cost of the synthesis. The power-law signals exhibit a lower number of total bound/generated proteins and energy costs compared with the other signal types. Variations with respect to the signal time $\tau_s$ and collision probability $p_{coll}$ are also reported in the middle and lower rows of Fig. \ref{f5} for a step signal function. As expected, all the total quantities increase with the signal time. The total number of generated proteins and energy cost $N_{tot}, E_{tot}$ do not depend on the collision probability in contrast to the total number of bound proteins. The total number of bound proteins is larger for higher collision probabilities specially for small $L_p$ where the probability of binding $p_{on}$ is greater than that of larger proteins.

\begin{figure}
\includegraphics[width=16cm]{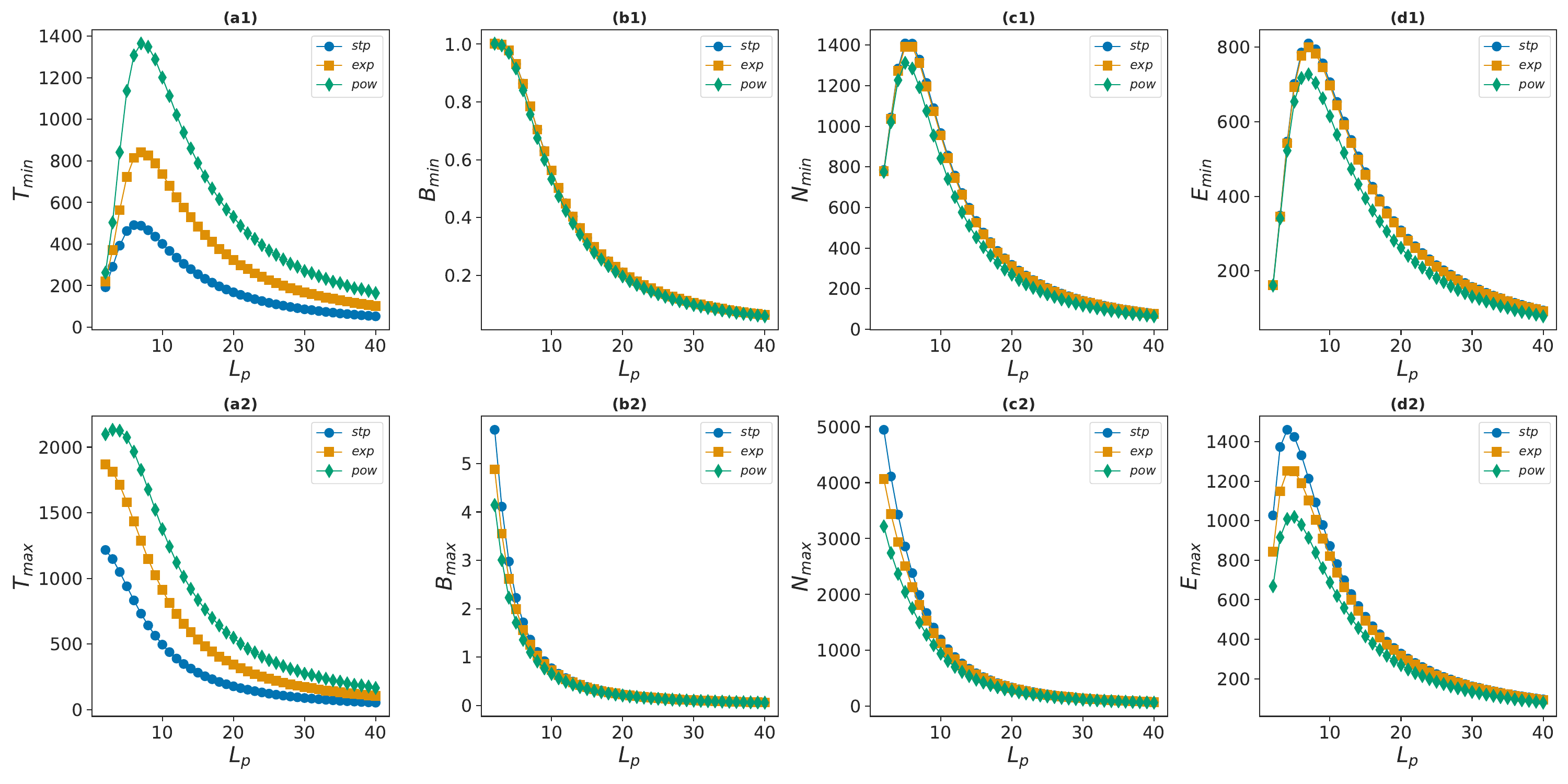} 
\caption{The min/max of the main quantities ($T_{min/max}, B_{min/max}, N_{min/max}$ and $E_{min/max}$) vs $L_p$ for the three types of signals $\phi(t)$: step (stp), exponential (exp), and power law (pow) functions. Here $L_c=1, L_d=20, \tau_s=4L_d^2, T=12\tau_s, \tau_k=1/72, p_{coll}=A_c/A_d, \lambda=1/4, \gamma=2, \delta=2, 1/z_{off}=0.1$. The cell and sphere surfaces are $A_{c,d}=4\pi L_{c,d}^2$. Top panels ((a1),(b1),(c1),(d1)) show the conditional mean values of the binding time, the number of bound proteins, and the number and energy cost of synthesized proteins when $B(t)$ becomes nonzero for the first time at $T_{min}$. Bottom panels ((a2),(b2),(c2),(d2)) show the same quantities when $B(t)$ is maximal at $T_{min}$. The results are averaged over $10^6$ realizations of the signaling process (synthesis, diffusion and degradation, binding and unbinding).}\label{f6}
\end{figure}

In Fig. \ref{f6} we see how the min/max quantities $T_{min,max}, B_{min,max}, N_{min,max}$ and $E_{min,max}$ change with the size of protein $L_p$ for different signals. The step signals display the lowest min/max time and at the same time the largest number of min/max protein numbers and energy costs. On the other hand the power-law signals show the smallest $N_{min,max}$ and $E_{min,max}$ at the expense of a smaller number of min/max bound proteins and larger $T_{min,max}$. In the Supplementary Information (SI Figs. 2,3), we report the changes in above quantities vs the signal time $\tau_s$ and collision probability $p_{coll}$ for a step signal function. The behavior of $B_{min}$ shows that the probability of binding in time period $T$ decreases monotonically with $L_p$ whereas $T_{min}, N_{min}$, and $E_{min}$ are smaller for very small or large protein lengths. Regarding the maximal quantities, we observe that $B_{max}$ and $N_{max}$ decrease monotonically with $L_p$, but $T_{max}$ and $E_{max}$ can exhibit a maximum for very small and intermediate values of $L_p$, respectively.

Based on the above quantities we define some measures of performances (efficiencies) to quantify the amount of transferred information per the consumed energy, the process time, and the number of synthesized proteins. We assume that all the information is encoded on the protein, and there is no information in e.g. when,
or whether, the protein arrives. Moreover, the information encoded on each
protein is independent of the others and each bound protein satisfies $\propto L_p$ bits of information. This means that the effective number of informative sequences of such a protein is $\propto e^{L_p}$. 

(i) the amount of received information per consumed energy at times $T_{min/max}$ and in total time $T$,
\begin{align}
\eta_E(T_{min})&=P(B_{tot}>0)\frac{L_pB_{min:B}}{E_{min:B}},\\
\eta_E(T_{max})&=P(B_{tot}>0)\frac{L_pB_{max:B}}{E_{max:B}},\\
\eta_E(T)&=\frac{L_pB_{tot}}{E_{tot}},
\end{align}

(ii) the amount of received information per spent time,
\begin{align}
\eta_T(T_{min})&=P(B_{tot}>0)\frac{L_pB_{min:B}}{T_{min:B}},\\
\eta_T(T_{max})&=P(B_{tot}>0)\frac{L_pB_{max:B}}{T_{max:B}},\\
\eta_T(T)&=\frac{L_pB_{tot}}{\tau_s},
\end{align}

(iii) the amount of received information per generated proteins,
\begin{align}
\eta_N(T_{min})&=P(B_{tot}>0)\frac{L_pB_{min:B}}{N_{min:B}},\\
\eta_N(T_{max})&=P(B_{tot}>0)\frac{L_pB_{max:B}}{N_{max:B}},\\
\eta_N(T)&=\frac{L_pB_{tot}}{N_{tot}}.
\end{align}

In each of the above measures we compare the amount of received information with the corresponding values of consumed energy, spent time, and number of generated proteins. Usually, the efficiency of information transmission is computed by the mutual information between the output of the source and input of the receiver \cite{BM-prl-2023,SAC-prr-2023} or capacity of the channel \cite{PJ-ieee-2016}. Instead, here we use measures of performances which are computationally easier to obtain than mutual information and still can capture the essence of an optimal communication process.

\begin{figure}
\includegraphics[width=16cm]{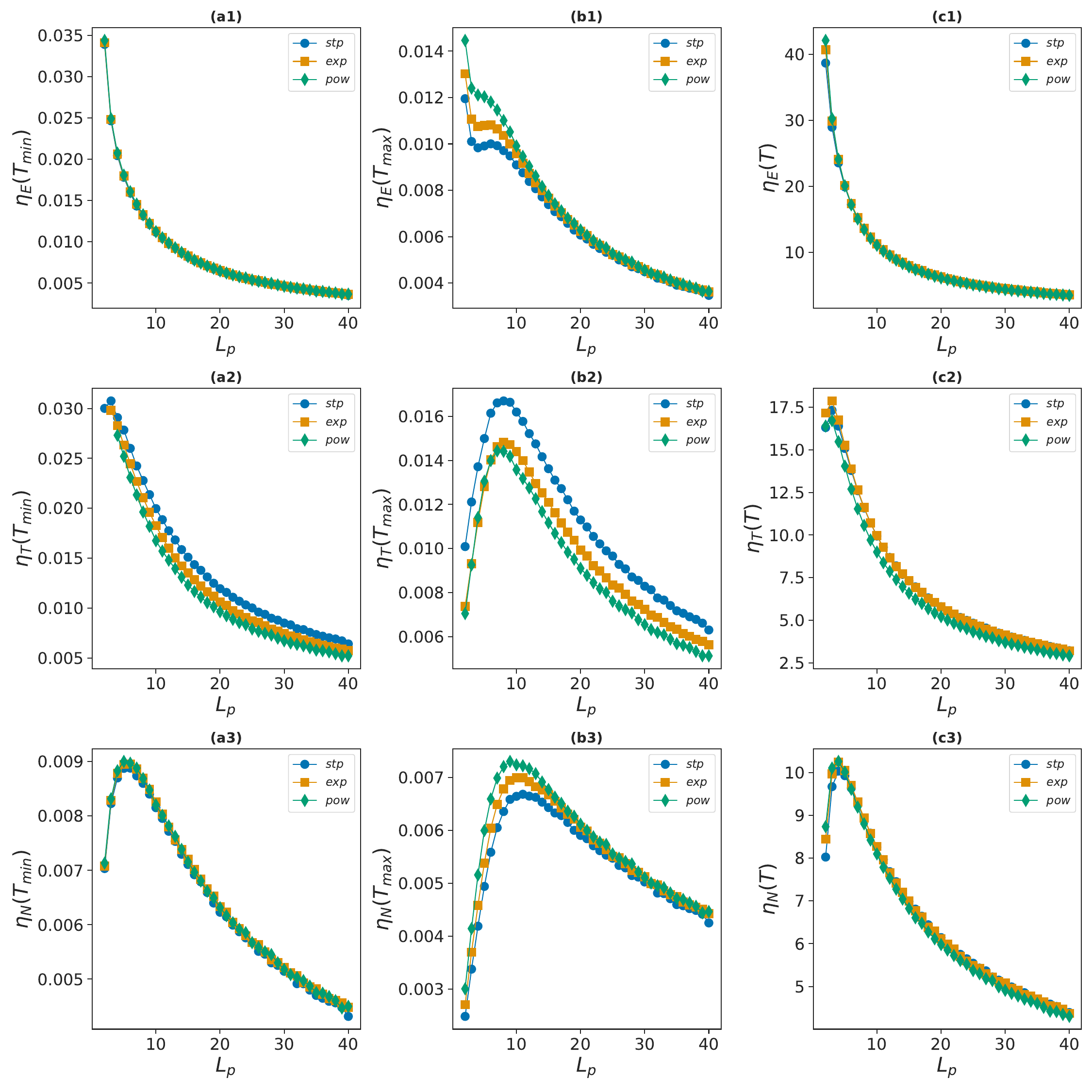} 
\caption{The information per energy $\eta_E$, per time $\eta_T$, and per number of generated proteins $\eta_N$ vs $L_p$ for the three types of signals $\phi(t)$: step (stp), exponential (exp), and power law (pow) functions. Here $L_c=1, L_d=20, \tau_s=4L_d^2, T=12\tau_s, \tau_k=1/72, p_{coll}=A_c/A_d, \lambda=1/4, \gamma=2, \delta=2, 1/z_{off}=0.1$. The cell and sphere surfaces are $A_{c,d}=4\pi L_{c,d}^2$. Panels ((a1),(b1),(c1)) show the conditional mean values of the $\eta_E$ efficiency at $T_{min/max}$ and the mean value of $\eta_E$ computed by the total quantities in time period $T$. Panels ((a2),(b2),(c2)) and ((a3),(b3),(c3)) show the $\eta_T$ and $\eta_N$ efficiencies for the same situations, respectively. The results are averaged over $10^6$ realizations of the signaling process (synthesis, diffusion and degradation, binding and unbinding).}\label{f7}
\end{figure}

Figure \ref{f7} shows how the above measures vary with the size of a protein for the three signals. The energy and number efficiencies $\eta_E, \eta_N$ are a bit larger for a power-law signal. The time performance of a step function is larger than that of the other signals in the min/max cases but not for the total one $\eta_T(T)$ and for small $L_p$. The results for different signal times and collision probabilities are given in the Supplementary Information (SI Figs. 4,5). First, we observe that a maximum in $\eta_E(T_{max})$ appears for larger $\tau_s$ and shifts to larger protein sizes by increasing $p_{coll}$. We do not see this effect in $\eta_E(T_{min})$ and $\eta_E(T)$. Second, the other efficiency measures $\eta_T(T_{min}), \eta_T(T_{max}), \eta_T(T)$ and $\eta_N(T_{min}), \eta_N(T_{max}), \eta_N(T)$ always display a maximum which is present at larger $L_p$ for larger signal time $\tau_s$ and larger collision probability $p_{coll}$. 

It should be mentioned that the qualitative behavior of the above observations are not very sensitive to the exact values of the model parameters. In the Supplementary Information, we report more results for other values of the free model parameters which support the main picture presented in this work. It is the case even for efficiency measures which are not proportional to $L_p$, that is the received information is independent of the messenger length, if we use a smaller diffusion coefficient $D=1/L_p^{0.33}$. The key point as stated before is to have degradation or binding mechanisms which compensate for the low synthesis rate and high energy cost of very large proteins and the smaller diffusion coefficient of larger proteins.

\section{Conclusion}  
We studied a model of chemical communication by diffusion to see how the messenger (protein) size affects the efficacy of information transfer from a central source (cell) to a number of binding sites at a fixed distance from the source. The whole process consisted of three stages to take into account the dependence on the synthesis rate and energy cost, the diffusion and degradation, and the binding of the proteins to the receptors. Accordingly, we defined some appropriate measures to quantify the performance of such a process regarding the energy costs, time scales, and the number of proteins on the binding sites. 
Note that our assumptions in definition of the model and performance measures may not exactly hold in some biological scenarios. But as we stated in the main text, here we suggest a working principle that is based on some simple efficiency measures and can exhibit the observed optimality of messenger size. 

The main observations from the numerical simulations are:
(I) The total efficiency measures $\eta_T(T)$ and $\eta_N(T)$ display a maximum around a protein length $L_p<10$. But $\eta_E(T)$ decreases monotonically with the protein size $L_p$. These behaviors do not change too much with the signal time and collision probability. (II) $\eta_N(T_{min})$ is the only efficiency measure computed at $T_{min}$ which displays a maximum about a small value of $L_p$. This optimal length increases with both $\tau_s$ and $p_{coll}$. (III) All measures $\eta_E(T_{max}), \eta_T(T_{max})$ and $\eta_N(T_{max})$ display a maximum at protein lengths $L_p>10$ which increases with the signal time and collision probability. $\eta_E(T_{max})$ is somehow special in that a local maximum appears which can become a global one by increasing $\tau_s$ and $p_{coll}$. In summary, we observed two different length scales for the messengers, corresponding to small and large $L_p$, depending on the relevance of the quantity ($E,T, N$) and the time scale $T_{min},T_{max},\tau_s$. 

Note that in this study we ignore interactions between the proteins. This is a reasonable approximation as long as the number and size of synthesized proteins is small compared with respect to the space of communication. Otherwise the interactions can play an important role in the performance of the process. Moreover, in this work we treat the messenger proteins as passive particles diffusing randomly due to the thermal noise of the environment. It would be interesting to consider also information communication by active particles which in addition to the thermal energy consume for example internal sources of energy to navigate in response to interactions with other chemicals which are present in the system.

\acknowledgments
This work was performed using the ALICE compute resources provided by Leiden University.


\end{document}